# Photocurrent Enhancement due to Spin-Exchange Carrier Multiplication in Films of Manganese-Doped 'Inverted' CdSe/HgSe Quantum Dots

Jungchul Noh, Clément Livache, Donghyo Hahm, Valerio Pinchetti, Ho Jin, Changjo Kim, and Victor I. Klimov*

Nanotechnology and Advanced Spectroscopy Team, C-PCS, Chemistry Division, Los Alamos National Laboratory, Los Alamos, New Mexico 87545, USA

*Corresponding author's e-mail: klimov@lanl.gov

**Incorporation of manganese impurities into II-VI semiconductors results in a dramatic change in their properties due to strong exchange interactions between the Mn ion and the semiconductor host. In colloidal quantum dots (QDs), these interactions result in a rapid bidirectional energy transfer between the magnetic impurity and the QD intrinsic states, which is characterized by an extremely high energy transfer rate of more than ~5 eV ps$^{-1}$. This rate is higher than the rate of energy loss due to phonon emission (typically, ~1 eV ps$^{-1}$ or less), so Mn-QD interactions could in principle be used to capture and utilize the kinetic energy of the hot carrier before it is lost to phonons. Here, we demonstrate that by using Mn-doped CdSe/HgSe core/shell QDs, we can efficiently convert the kinetic energy of a hot exciton into an additional electron-hole pair (exciton). This carrier multiplication process occurs through rapid capture of a hot exciton by a Mn ion, which then undergoes spin flip relaxation, producing two excitons near the QD band edge. Due to the inverted geometry of the CdSe/HgSe QDs, both electrons and holes resulting from carrier multiplication occupy the QD shell, allowing them to be easily extracted from the QD for potential use in electro-optical devices or chemical reactions.**

In the presence of carrier multiplication (CM), one absorbed photon produces not one but several electron-hole (e-h) pairs (excitons)[1,2]. As a result, the internal quantum efficiency (IQE) of the photon to e-h pair conversion ($\eta_{eh}$) becomes greater than unity (or 100%). CM can potentially benefit photovoltaic (PV), and photodetection devices[2-5]. It can also be useful in photochemistry[6,7], especially in the case of chemical transformations requiring multiple reduction/oxidation steps. CM occurs in both extended (bulk) and low-dimensional semiconductor systems. This process has been studied particularly thoroughly for colloidal nanocrystals of various morphologies[3,8-11] including quantum dots (QDs), nanorods, and nanoplatelets. The advantage of these materials is the reduction of the CM threshold ($E_{th,CM}$), resulting from the relaxation of translational momentum conservation[8,9,12]. In particular, using properly designed heterostructures, such as PbSe/CdSe core/shell QDs[13] or PbS/CdS Janus particles,[14] it was possible to reduce $E_{th,CM}$ to two bandgaps ($2E_g$), which corresponds to the ultimate limit determined by the energy conservation law.

CM is usually explained by impact ionization, during which a valence-band electron is excited into the conduction band due to a Coulomb collision with an energetic (hot) electron or hole[1,2] (Fig. 1a). A process competing with impact ionization is hot-carrier cooling via phonon emission. For conventional ('direct') Coulombic interactions, the energy-loss rate due to phonons ($r_{loss} = E_{phon}/\tau_{phon}$; $E_{phon}$ and $\tau_{phon}$ are the effective phonon energy and emission time, respectively) is at least a factor of 3 higher than the energy-gain rate due to Coulomb collisions ($r_{gain} = E_g/\tau_{ii}$, $\tau_{ii}$ is the characteristic impact-ionization time).[15-17] The unfavorable ratio between $r_{gain}$ and $r_{loss}$ ($r_{gain}/r_{loss} < 0.3$) is the main factor limiting the CM yield (defined as $\eta_{CM} = \eta_{eh} - 1$) in both bulk and low-dimensional semiconductors.

Recent studies of manganese-doped QDs demonstrate that $r_{gain}$ can be dramatically enhanced by employing not 'direct' but 'exchange' carrier-carrier Coulomb interactions mediated by a magnetic impurity[15,18]. In particular, using exchange interactions, it was possible to achieve very fast (sub-picosecond) excitation transfer between a Mn dopant and a QD. During this process, the Mn ion undergoes a spin-flip transition from the excited $^4T_1$ state to a ground $^6A_1$ state, accompanied by the change of the total ion's spin by $\Delta S = 1$. Because of spin-conservation, this transition couples only to a spin-1 'dark' QD exciton state. Hence the Mn-to-QD (or QD-to-Mn) excitation transfer occurs as energy-conserving spin exchange. Notably, the primary driving force of this transfer is 'exchange' Coulomb interactions as coupling between the Mn spin-flip transition and QD excitonic states due to 'direct' Coulomb interactions is negligibly small[19,20]. Based on these considerations, we will refer to processes involving QD-Mn energy exchange as 'spin-exchange' (SE) processes.

Due to a large energy of the Mn transition ($E_{Mn} = 2.1$ eV) and very short time constants of Mn-QD energy exchange ($\tau_{SE} = 100$ to $300$ fs), the corresponding energy gain rate (estimated from $E_{Mn}/\tau_{SE}$) reaches very large values (>5 eV ps$^{-1}$) that exceed $r_{loss}$ due to photon emission (typically, ~1 eV ps$^{-1}$ or less[15,18]). Thus, instead of trailing $r_{loss}$ by a factor of 3 or more for 'direct' Coulomb interactions, $r_{gain}$ becomes greater than $r_{loss}$ by a factor of at least 5. This opens up interesting possibilities for energy interconversion, especially in schemes aimed at using the kinetic energy of unrelaxed, hot carriers.

Previously, the favorable energy gain/loss ratio in Mn-doped QDs was used to demonstrate highly efficient photoemission induced by visible light pulses[18]. This process was realized as a two-step Auger re-excitation, where a band-edge electron was excited to the vacuum state outside the QD via two successive energy transfers from excited Mn ions. In undoped QDs, the second step of Auger re-excitation would have been hampered by the relaxation of the hot electrons through phonon emission. However, in Mn-doped structures, due to the extremely high SE energy transfer rates, the hot electron was efficiently excited to the vacuum state before losing any significant energy to phonons.

More recently, QD-Mn interactions were exploited to realize highly efficient SE-CM in Mn-doped PbSe/CdSe QDs, in which a lower-bandgap PbSe core was enclosed in a higher-bandgap CdSe shell[21]. In these structures, SE-CM occurred in two steps (Fig. 1b): (1) SE energy transfer from a hot exciton delocalized throughout the QD to the interfacial Mn ion, followed by (2) energy- and spin-conserving relaxation of the excited Mn ion to generate two excitons (spin-0 'bright' and spin-1 'dark') in the PbSe core. Due to extremely short SE time scales, both SE steps occurred without considerable interference from phonon emission, resulting in high SE-CM efficiency.

To facilitate the utilization of carriers generated by SE-CM, in this study we synthesized Mn-doped 'inverted' CdSe(core)/HgSe(shell) QDs, in which the lower bandgap material is located in the shell region, making SE-CM-generated electrons and holes easily electrically accessible (Fig. 1c). This is an advantage over PbSe/CdSe QDs, in which charge extraction/transfer is hindered by the large potential barrier created by the CdSe shell. By varying the dimensions of the core/shell CdSe/HgSe structure, we tuned the QD bandgap to be slightly smaller than $E_{Mn}/2$ (= 1.05 eV), which is optimal for SE-CM. To detect the SE-CM and quantify its efficiency, we used a femtosecond transient absorption (TA) experiment to monitor the signatures of biexciton Auger decay. TA measurements revealed highly efficient SE-CM manifested by a sharp stepwise increase in IQE above 100% with increasing photon energy ($E_p$). In particular, $\eta_{eh}$ reached 157% at $E_p$ = 2.4 eV, which was only slightly above the SE-CM threshold determined by the Mn spin-flip transition energy (2.1 eV). The corresponding CM yield is 57% which is approximately 4 times higher than that of the reference undoped CdSe/HgSe QDs. Importantly, we also observed the SE-CM-related photocurrent enhancement in the QD films. The CM yields derived from the photocurrent measurements of Mn-doped compared to undoped QDs are consistent with those obtained from TA measurements. This indicates the potential usefulness of our Mn-doped CdSe/HgSe QDs in practical CM-enhanced photoconversion.

## QD synthesis and doping with manganese

We started our work on QD synthesis by optimizing undoped core/shell CdSe/HgSe/ZnS QDs. Bulk HgSe has an inverted band structure with a negative bandgap of –0.24 eV (ref [22]). In the case of quantum-confined nanostructures, this material develops a standard positive bandgap that can be tuned to energies of interest in the present study (~1 eV or less) by controlling the nanostructure size, which in our case is the thickness ($H_{HgSe}$) of the HgSe layer.

To prepare CdSe cores we use a procedure similar to that previously described in ref [23]. The synthesized CdSe particles are reacted with Hg and Se precursors ($HgCl_2$ and TOP-Se, respectively; TOP-Se is trioctylphosphine selenium; Fig. 2a). Due to the nearly identical lattice constants of CdSe and HgSe ($a_{HgSe}$ = 6.08 Å, $a_{CdSe}$ = 6.05 Å; ref [24]), the deposition of HgSe on top of CdSe cores occurs epitaxially, resulting in the formation of a spherically symmetric shell whose thickness ($H_{HgSe}$) is quite uniform over the entire ensemble of QDs (Fig. 2b). In the final step, the core/shell CdSe/HgSe QDs are coated with a thin protective layer ZnS by their reaction with $Zn(Ac)_2$ (zinc acetate) and ODE-S (octadecene-sulfur).

In Fig. 2b, we show transmission electron microscopy (TEM) images of the synthesized materials (CdSe cores and CdSe/HgSe core/shell QDs). In this example, the average core radius ($r_{CdSe}$) is 2.3 nm and the average total radius ($R_{CdSe/HgSe}$) is 2.6 nm. Based on these dimensions, the HgSe shell thickness is 0.3 nm, which corresponds to approximately 1 HgSe monolayer (ML). The synthesized structures exhibit excellent monodispersity. The standard deviation of the size for the core sample ($\Delta_r$) is ~0.11 nm (Supplementary Fig. 1a). It is virtually unchanged in the case of the core/shell QDs ($\Delta_R$ is ~0.13 nm; Supplementary Fig. 1b), suggesting an extremely high uniformity of the HgSe layer thickness across the QD ensemble.

In Fig. 2c, we show the approximate band diagram of the CdSe/HgSe/ZnS QD. Due to the large bandgaps of CdSe (1.75 eV) and ZnS (3.6 eV), the electron and hole wavefunctions (red and blue lines, respectively) are tightly confined within the HgSe layer. As a result, the band-edge optical transition energy (nominal bandgap of the QD) is expected to be determined primarily by $H_{HgSe}$, with a weaker influence from $R_{CdSe}$.

In this work, we will use the peak photoluminescence (PL) spectral energy as a measure of the bandgap of the QD samples. The expected strong dependence of $E_g$ on the HgSe layer thickness is evident from Fig. 2d which shows PL spectra of a series of samples with a fixed CdSe core radius ($r_{CdSe}$ = 2.3 nm) and $H_{HgSe}$ varying from 0.3 to 1.9 nm (~1 to ~6 HgSe MLs). All samples contain a final monolayer of wide-bandgap ZnS to enhance the stability of the QDs. Based on the PL spectra, $E_g$ is around 1.01 eV for $H_{HgSe}$ of 1 ML and it drops to 0.61 eV for the 6 ML sample. Although all these samples meet the energy conservation requirement for SE-CM ($E_g < E_{Mn}/2$ = 1.05 eV), here we focus on the thinner shell samples (~ 3 MLs) since they can maximize the potential benefits of SE-CM by providing higher PV efficiency (see the section "Implications for solar energy conversion").

To prepare Mn-doped samples, we apply a diffusion doping procedure to CdSe cores[25,26] and then proceed with the same synthesis steps as for undoped samples. In Fig. 3a, we show example TEM images and optical spectra of undoped and Mn-doped CdSe cores prepared using doping time $t_{Mn}$ = 15 min (Methods). After doping, the core radius increases slightly (from 2.3 to 2.4 nm). Simultaneously, the band-edge (1S) absorption peak experiences a blue shift (from 2.1 to 2.2 eV). These observations suggest that $Mn^{2+}$ is introduced into the QD not as a MnSe shell (which would redshift the 1S peak) but as a substitutional impurity that replaces part of the $Cd^{2+}$ ions in the QD[25,26]. The resulting formation of $Mn_xCd_{1-x}Se$ alloy increases the effective bandgap of the QDs, since $E_g$ of MnSe (=2.9 eV) is larger than $E_g$ of CdSe (=1.75 eV). The increase in $t_{Mn}$ results in a progressively stronger blue shift of the 1S peak, as expected for the increasing Mn content ($x$) in the sample (Supplementary Fig. 2). This assessment is supported by elemental analysis using inductively coupled plasma atomic emission spectroscopy (ICP-AES). According to ICP-AES measurements (Supplementary Table 1), the Mn content increases from ~6% (of total cations) for $t_{Mn}$ = 1 min to ~11% for $t_{Mn}$ = 15 min.

TEM and optical characterization of the Mn:CdSe/HgSe/ZnS QD sample ($t_{Mn}$ = 15 min) and the reference undoped QDs with similar dimensions are shown in Fig. 3b. The final radius of the Mn:CdSe/HgSe/ZnS QDs is 3.75 nm with $H_{HgSe}$ of ~1 nm (~3 HgSe MLs) and $H_{ZnS}$ of ~0.4 nm (~1 ZnS ML). Importantly, the deposition of the shell layers does not have a significant effect on the Mn content in the QDs. In particular, in the Mn:CdSe/HgSe/ZnS QD sample prepared using $t_{Mn}$ = 15 min, the Mn content is 2.3% of all cations (Supplementary Table 1). However, when evaluated relative to the total amount of Cd and Mn ions, the Mn content is 9.2% which is very close to that in the original CdSe cores ($x$ = 10.8%).

To determine whether the Mn dopants are coupled to the HgSe shell via exchange interactions (a prerequisite for the implementation of SE-CM), we performed magnetic circular dichroism (MCD) measurements of Mn:CdSe/HgSe/ZnS QDs at the NIR band edge which is due to the light absorption in the HgSe shell (Methods). The measured MCD signal ($\Delta A_{MCD}$, a quantity defined by the difference between absorption coefficients for left- and right-handed circularly polarized light) exhibits a temperature- and magnetic-field-dependent behavior, consistent with an exchange interaction of the embedded spin-5/2 Mn ions with excitons localized in the HgSe-shell (Supplementary Fig. 3). In particular, $\Delta A_{MCD}$ increases with decreasing temperature due to the increase of the internal field created by the Mn spins, an increasing number of which are aligned by the external field. A decrease in temperature leads to the appearance of a peculiar dependence of $\Delta A_{MCD}$ on the magnetic field, which shows saturation in the high-field region (Supplementary Fig. 3). This means reaching a regime in which the internal field is maximized after all Mn spins are completely aligned by the applied field.

By fitting the measured field-dependent $\Delta A_{MCD}$ to the Brillouin function (Supplementary Fig. 3), we find that the effective concentration of Mn paramagnetic spins that interact with the shell-localized exciton ($x_{eff}$) is 0.8% of all cations 'sampled' by the exciton wavefunction. Considering that the percentage of Hg ions in the measured sample is ~48% (Supplementary Table 1), this corresponds to a total Mn content of ~0.4% with respect to all cations, or approximately one sixth of all Mn ions in the QDs as determined by ICP-AES. Thus, a significant fraction of the Mn ions introduced by the doping procedure are exchange coupled with shell-localized excitons and can thus mediate the SE-CM process.

Based on the PL spectra of undoped and Mn-doped CdSe/HgSe/ZnS QDs (Fig. 3b), their bandgaps are 0.81 and 0.92 eV, respectively. The latter value is only slightly smaller than $E_{Mn}/2$, which should allow us to maximize the SE-CM efficiency by minimizing the energy loss in the second SE step, where the energy of the excited Mn ion is converted into 2 near-band-edge excitons. Since the Mn spin is 'flipped' in this process from the 3/2 to the 5/2 state, the total spin of the excitons created by SE-CM must be equal to 1 which can be achieved if one of them is 'bright' (spin 1) and the other is 'dark' (spin 0)[21]; Fig. 1c.

**Transient absorption CM measurements**

To detect CM and quantify its efficiency, we monitor the population of the band-edge QD states using a femtosecond TA experiment in which the change in the 1S absorption signal ($\Delta\alpha_{1S}$) induced by the pump pulse is probed by a broadband femtosecond white light continuum pulse (Methods). The presence of photoinjected carriers in the band edge states results in 1S bleaching ($\Delta\alpha_{1S} < 0$) caused by state filling[27]. The magnitude of $\Delta\alpha_{1S}$ is determined primarily by the occupation factor of the $1S_e$ electron level[28]. In II-VI QDs, the $1S_e$ state is two-fold spin-degenerate. Therefore, $\Delta\alpha_{1S}$ scales linearly with the total number of e-h pairs per-dot ($N_{eh}$) until $N_{eh}$ reaches 2. At higher $N_{eh}$, $\Delta\alpha_{1S}$ saturates because if the number of electrons exceeds 2, the extra electrons are forced to go to the higher energy $1P_e$ state[27].

A distinctive signature of CM in TA measurements is the fast component of the 1S bleach decay associated with the Auger recombination of multiexcitons[8]. In this process, an e-h pair recombines, transferring its energy to a third carrier (electron or hole) located in the same QD. In strongly confined nanocrystals, it is characterized by extremely fast time constants (the biexciton Auger lifetime, $\tau_{A,XX}$, is typically between 10s to 100s picoseconds), which are much shorter than those of a single-exciton decay[29,30]. They are also significantly shorter than the biexciton radiative time constant ($\tau_{R,XX}$). Therefore, the total biexciton lifetime ($\tau_{XX}$) is defined by $\tau_{A,XX}$. If a fast Auger decay component is detected when the average number of photons absorbed per QD per pulse is much less than 1 ($\langle N_{abs}\rangle \ll 1$), this means that multiexcitons are generated via the CM[8].

To quantify the biexciton lifetimes, we performed TA measurements of reference (undoped) and Mn-doped core/shell samples using excitation with low-energy photon ($E_p$ = 1.2 eV) for which CM is impossible because $E_p < 2E_g \approx 1.8$ eV. In Fig. 4a, we show a series of tail-normalized 1S bleaching dynamics of the undoped sample recorded for progressively increasing per-pule pump fluences ($j_p$) ranging from 0.065 to 1.30 mJ cm$^{-2}$. Based on the absorption cross-sections at 1.2 eV of $\sigma_{abs} = 5.6 \times 10^{-16}$ (Methods), these fluences correspond to $\langle N_{abs}\rangle = (\sigma_{abs}\, j_p)/E_p$ from 0.19 to 3.8.

At sub-single-exciton pump levels ($\langle N_{abs}\rangle < 1$), the TA traces of the undoped sample are nearly flat on the time scale of TA measurements ($\Delta t < 100$ ps), reflecting the slow decay of the single-exciton states. At $\langle N_{abs}\rangle > 1$, the TA dynamics develops a fast component due to the Auger decay of multiexcitons generated by successive absorption of photons from the same pulse. Subtracting the tail-normalized single-exciton time transient measured at $\langle N_{abs}\rangle = 0.19$ from the higher pump intensity traces ($\langle N_{abs}\rangle = 0.38, 0.95, 1.90$, and 3.80), we isolate the biexciton decay, yielding $\tau_{A,XX}$ of 16 ps (inset in Fig. 4a).

The dynamics of 1S bleaching of the Mn-doped sample at low-pump-levels show a behaviour that is more complex than that of the reference sample (Fig. 4b). In addition to the slow component they exhibit an initial 'fast' signal, probably related to electron trapping at defect sites introduced during the doping procedure. The time constant of the fast component ($\tau_f$) is ~40 ps. This is appreciably longer than the expected biexciton Auger lifetime, suggesting the applicability of the 'subtraction' procedure to isolate the biexciton decay. Fitting the dynamics obtained with this procedure to a single-exponential relaxation, we obtain $\tau_{A,XX}$ of 17 ps (Fig. 4b, inset), which is similar to $\tau_{A,XX}$ obtained for the undoped QDs.

To make CM energetically possible, we change the pump-photon energy to $E_p = 2.4$ eV. This is larger than either $2E_g$ or $E_{Mn}$, which opens up hot-exciton relaxation pathways via both conventional CM and SE-CM. The former is expected to be active in both undoped and Mn-doped samples, whereas the latter is only active in Mn-doped QDs. The existence of CM in the undoped sample is evident in Fig. 4c, where we show two tail-normalized low-pump-intensity TA traces ($\langle N_{abs} \rangle$ = 0.38) obtained with excitation below ($E_p$ = 1.2 eV, black line) and above ($E_p$ = 2.4 eV, blue line) the nominal CM threshold $2E_g$. The trace recorded below the CM onset shows primarily a slowly relaxing single-exciton signal. In contrast, the dynamics obtained using 2.4-eV pump photons exhibits a clearly discernible fast component whose time constant (16 ps) is consistent with the Auger decay of a biexciton. This is a typical signature of CM, where biexcitons are generated by single absorbed photons at low (nominally sub-single exciton) pump levels.

To quantify the CM yield, we assume that the exciton and biexciton generation probabilities per absorbed photon are $q_1$ and $q_2$, respectively, and that they are normalized such that $q_1 + q_2 = 1$. Using these quantities, we can present the photon-to-exciton conversion efficiency as $\eta_{eh} = q_1 + 2q_2$. We can further express the amplitude of the early time 1S signal ($A$) before Auger decay (pump-probe delay $\Delta t < \tau_{A,XX}$) as $A = \gamma(q_1 + 2q_2)$, where $\gamma$ is a proportionality constant. After Auger decay, the initially generated biexcitons are converted to single excitons. Therefore, the late-time TA signal observed at $\tau_{A,XX} < \Delta t < \tau_X$ ($\tau_X$ is the single-exciton time constant) is given by $B = \gamma(q_1 + q_2) = \gamma$. From these expressions, $A/B = (q_1 + 2q_2) = \eta_{eh}$ and $\eta_{CM} = \eta_{eh} - 1 = A/B - 1$ (Fig. 4d). Thus, the CM yield can be directly derived from the $A/B$ ratio measured at low pump fluences, when $\langle N_{abs} \rangle \ll 1$ (refs [8,31]). This method is implemented in Fig. 4c (inset, see also Supplementary Fig. 4a), where we plot the ratio of the TA signals before and after Auger decay as a function $\langle N_{abs} \rangle$. By extrapolating the plotted dependence to $\langle N_{abs} \rangle = 0$, we obtain $\eta_{eh}$ of 113%, which yields $\eta_{CM}$ = 13%.

Next, we analyze the effect of CM on the TA dynamics of the Mn-doped QD sample. In Fig. 4e, we show the 1S-bleaching relaxation of this sample obtained using excitation with 1.2 eV (black line) and 2.4 eV (red line) photons at low excitation fluences for which $\langle N_{abs} \rangle$ is smaller than 0.1.

As mentioned earlier, the 1.2 eV trace shows a rather large initial short-lived component due to the defect-related single-exciton recombination. An even faster component appears in the regime where CM is possible ($E_p$ = 2.4 eV). Its time constant (13 ps) is close to the biexciton Auger lifetime determined earlier (Fig. 4b and Supplementary Fig. 4b), indicating that it is due to CM.

In order to quantify the CM yield for the Mn-doped sample, we need to modify the procedure illustrated in Fig. 4d to account for the fast single-exciton decay component. For this purpose, we assume that the fraction of QDs ($f_f$) in the QD ensemble undergoes fast nonradiative decay with a time constant ($\tau_{f,X}$) that is still slower than $\tau_{A,XX}$ but faster than the time scale of the TA measurements (~1 ns in our case). The other part of the QD ensemble ($f_s = 1 - f_f$) is assumed to exhibit a slower decay with a time constant much larger than $\tau_{A,XX}$. As suggested in ref [32], one can take into account the inhomogeneity of single exciton relaxation in a QD sample by renormalizing the $A/B$ ratio measured in the presence of CM ($A_1/B_1$ observed for $E_p > E_{th,CM}$) based on TA measurements in the absence of CM ($A_2/B_2$ observed for $E_p < E_{th,CM}$). In the case of CM, $A_1$ is contributed by both 'fast' and 'slow' QD subensembles, whereas only QDs with 'slow' dynamics contribute to $B_1$. This yields $A_1 = \gamma(f_s + f_f)(q_1 + 2q_2) = \gamma(q_1 + 2q_2)$, $B_1 = \gamma f_s(q_1 + q_2)$, and $A_1/B_1 = Q/f_s$. In the case without CM, when $q_2 = 0$ and $q_1 = 1$, $A_2 = \gamma$, $B_2 = \gamma f_s$ and $A_2/B_2 = 1/f_s$. Therefore, $\eta_{eh}$ can be found from $\eta_{eh} = (A_1/B_1)/(A_2/B_2)$.

In practice, the above procedure can be implemented by plotting the tail-normalized 1S-bleaching dynamics measured at $\langle N_{abs} \rangle \ll 1$ below and above the CM threshold, from which the $A_1/A_2$ ratio can be determined and then used instead of the $A/B$ ratio in the CM efficiency analysis (Fig. 4f). From an analysis of the $A_1/A_2$ ratio for the Mn-doped sample plotted as a function of $\langle N_{abs} \rangle$ (Fig. 4e, inset), we obtain $\eta_{eh}$ = 157%, which yields $\eta_{CM}$ = 57%. The latter value is more than 4 times higher than $\eta_{CM}$ for the undoped sample for the same photon energy, indicating that the observed CM process is dominated by the Mn ion-mediated SE pathway, as shown in Fig. 1c.

Based on TA measurements with a spectrally tuneable excitation source, the multiexciton yield of the Mn-doped sample shows a sharp stepwise increase above the nominal SE-CM threshold $E_{Mn}$ (solid red circles in Fig. 5a and Supplementary Fig. 5), which is a prerequisite for obtaining a strong impact of SE-CM on practical solar energy conversion (see the section "Implications for solar energy conversion"). In particular, $\eta_{CM}$ reaches almost half of the ideal 100% value only 0.1 eV above $E_{Mn}$ ($\eta_{CM}$ = 46% at $E_p$ = 2.2 eV).

## Photocurrent measurements

An important advantage of the new Mn-doped CdSe/HgSe QDs compared to previously realized Mn-doped PbSe/CdSe structures is their 'inverted' geometry, which provides easy electrical access to both the electron and the hole and is therefore well suited for use in electro-optical devices such as solar cells and photodetectors. To evaluate the potential of our inverted QDs in such applications, we performed preliminary studies of their photoconductive response using densely packed solid-state films made from QDs stripped of their original long surface ligands to improve interdot coupling (Supplementary Figs. 6,7). In these experiments, we monitored the bias-dependent photocurrent using a spectrally tuneable continuous wave excitation source. From the ratio of the measured photocurrent to the absorbed photon flux, we determine the quantity $\eta'_{cur}$ which is proportional to the absorbed photon to current conversion efficiency ($\eta_{cur}$). We then normalize $\eta'_{cur}$ to be equal to 1 (or 100%) at photon energies below the energy-conservation-defined CM threshold of $2E_g$ and find the nominal CM yield from $\eta_{CM,cur} = \eta'_{cur} - 1$.

The data obtained by the above procedure for the undoped and Mn-doped QDs are shown in Fig. 5a (open and solid triangles, respectively) together with $\eta_{CM}$ obtained from TA measurements. For the doped sample, we observe sharp growth of $\eta_{CM,cur}$ at energies above ~2 eV, after which $\eta_{CM,cur}$ reaches 46% at 2.34 eV and eventually 54% at 2.95 eV. The undoped sample also shows the increase in $\eta_{CM,cur}$. However, it is much weaker and $\eta_{CM,cur}$ is only ~7% at 2.95 eV. Importantly, both the spectral dependence of $\eta_{CM,cur}$ and its absolute values are in close agreement with the data obtained from TA measurements (Fig. 5a, black and red circles for the undoped and doped QDs, respectively). This suggests that the enhanced production of e-h pairs obtained by SE-CM leads to enhanced photocurrent in a practical photoconductive device.

An important question regarding the photocurrent measurements is why the CM-related photocurrent enhancement is not suppressed by the fast Auger recombination of biexcitons produced by SE-CM. Although we cannot provide a definitive answer to this question at present, we have a plausible explanation for our observations. In particular, previous research into a photoresponse of QD films shows that the photoconductivity of QDs is often dominated by one type of carrier (electrons or holes), which has a higher mobility [33,34]. For example, previous studies of PbSe QD films revealed that their photoconductivity was almost exclusively due to mobile band-edge holes, as photogenerated electrons were quickly captured by weakly coupled intragap states (presumably attributed to surface sites) [33]. A similar single-carrier transport mechanism was also indicated in studies of PbS QD-based optical field-effect transistors, in which intragap electron traps manifested themselves in the spectrally resolved photocurrent as a distinct band below the QD band edge [34].

Such single-carrier photoconductivity is likely to be realized in our inverted QD films. In this scenario, one type of photogenerated carrier is rapidly trapped in intragap states, which should dramatically slow down Auger or other recombination due to the reduced overlap of localized intragap and extended band-edge QD states. This increases the time available for an untrapped carrier to escape from the initially excited QD and thereby avoid Auger recombination if the initially generated state was a biexciton.

**Implications for solar energy conversion**

To elucidate the potential implications of SE-CM for solar photoconversion, we model its impact on the power conversion efficiency (PCE) of solar cells using theoretical approaches of refs [4,5]. In our simulations, we approximate the spectral dependence of CM yield using linear interpolation of TA measurements and assuming that $\eta_{CM}$ is 0 below $E_{Mn}$ = 2.1 eV and is 'flat' (= 164%) above 3.6 eV (the highest spectral energy used in TA experiments). The CM-induced enhancement of PCE ($\Delta\eta_{PCE} = \eta_{PCE,CM} - \eta_{PCE}$, where $\eta_{PCE,CM}$ and $\eta_{PCE}$ are PCEs with and without CM, respectively) is primarily due to the increase in photocurrent. It can be assumed that this is due to the effective increase in the solar flux density ($\Delta\phi_s$) at energies above the CM threshold. $\Delta\phi_s$ can be related to $\eta_{CM}$ and the original solar flux density ($\phi_s$) by the relation $\Delta\phi_s = \int_0^\infty \eta_{CM}(v)\varphi_s(v)\,dv$, where $v$ is the photon frequency, and $\varphi_s(v)$ is the spectral distribution of $\phi_s$.

In Fig. 5b we show the spectral dependence of $\Delta\phi_s$ calculated for the experimental CM yields. From this data we obtain that the CM-induced flux enhancement normalized by the total solar flux ($\delta_s = \Delta\phi_s/\phi_s$) is 9.3%. Neglecting the weak effect of CM on the open-circuit voltage, this directly translates into the PCE enhancement ($\delta_{PCE} = \Delta\eta_{PCE}/\eta_{PCE} = \delta_s$), which is also expected to increase by 9.3%. This gives an absolute PCE value of $\eta_{PCE,CM}$ = 35% if we use $\eta_{PCE}$ = 32% calculated using the standard detailed-balance approach for $E_g$ = 0.92 eV, which is equal to the bandgap of our Mn-doped QDs.

To evaluate the maximum PCE enhancement achievable with SE-CM, we performed PCE simulations for different QD bandgaps assuming ideal SE-CM yields ($\eta_{CM}$ = 1 for photon energies above $E_{Mn}$ = 2.1 eV and zero below it). The results of these calculations are shown in Fig. 5c (red line). As expected, for bandgaps above $E_{Mn}/2$ = 1.05 eV, the calculated PCEs overlap with those for the no-CM case (black line). We observe a sharp spike in PCE at $E_g = E_{Mn}/2$, which corresponds to the optimal bandgap in the SE-CM regime. The corresponding PCE is 41.0%, which is a ~27% enhancement compared to the no-CM case. Interestingly, the peak PCE achievable with SE-CM is close the ultimate PCE limit (44.4%) calculated for the ideal 'stair-case-like' CM yield where $\eta_{CM}$ increases by 100% for each photon energy increment of $E_g$.

## Conclusions

To conclude, we have demonstrated that Mn-doped CdSe/HgSe/ZnS QDs exhibit a significant (more than 4-fold) enhancement in CM yield compared to reference undoped QDs with nominally identical structure. This enhancement occurs due to the SE-CM pathway mediated by Mn dopants. SE-CM occurs via rapid trapping of a hot photogenerated QD exciton by the Mn dopant, followed by spin-flip relaxation of the excited Mn ion to create two excitons near the band edge of the QD. The absorbed photon-to-exciton conversion efficiency, obtained using femtosecond TA measurements, increases sharply above the Mn spin-flip transition energy (~2.1 eV), reaching 146% at $E_p$ = 2.2 eV. Reference undoped QDs show $\eta_{eh}$ of 111% at the same photon energy.

Due to the 'inverted' geometry of CdSe/HgSe/ZnS QDs, both electrons and holes generated by CM are localized in the HgSe shell, making them easily accessible electrically. Indeed, close packed films of these QDs show characteristics of a well-behaved photoconductor. Interestingly, the efficiency of absorbed photon to current conversion increases sharply above the Mn spin-flip transition energy, and its spectral dependence closely correlates with the IQE dependence found in TA studies. This suggests that SE-CM impacts photocurrent and hence can be used in photoelectric devices.

PCE simulations of PV devices show that under optimal conditions, SE-CM can lead to a PCE of 41%, which is close to the theoretical limit (~44%) for a fully optimized conventional CM. In addition to PVs the SE-CM process can be useful in photochemistry, especially in multi-carrier reactions that can greatly benefit from the pairwise production of electrons and holes co-localized in time and space.

**Acknowledgements**

Spectroscopic and photocurrent-based studies of QDs and PCE modeling of PV devices in the presence of SE-CM were supported by the Solar Photochemistry Program of the Chemical Sciences, Biosciences and Geosciences Division, Office of Basic Energy Sciences, Office of Science, U.S. Department of Energy (DOE). The synthesis of the QDs and their microstructural characterization were supported by the Laboratory Directed Research and Development (LDRD) program at Los Alamos National Laboratory under project 20230275ER.

**Author contributions**

V.I.K. conceived the idea and coordinated the overall research efforts. J.N. synthesized undoped and Mn-doped QDs, obtained their TEM images, and measured their optical absorption and PL spectra. J.N., C.L., and V.P. performed the TA measurements. J.N. and V.P. performed the MCD measurements. J.N. and D.H. conducted simulation of power conversion efficiencies. J.N. and H.J conducted elemental analysis studies of the QDs. J.N. and C.K. performed photocurrent measurements. J.N. and V.I.K wrote the manuscript with contributions from other co-authors.

**Competing interests**

The authors declare no competing interests.

**Figure Legends**

Fig. 1| Conventional CM and SE-CM.

Fig. 2 | Synthesis of undoped CdSe/HgSe QDs with controlled shell thickness.

Fig. 3 | Structural characteristics and optical spectra of undoped and Mn-doped CdSe and CdSe/HgSe/ZnS QDs.

Fig. 4 | TA measurements of CM in CdSe/HgSe/ZnS and Mn:CdSe/HgSe/ZnS QDs.

Fig. 5 | CM yields obtained from TA and photocurrent measurements and implications of SE-CM for solar energy conversion.

## Methods

**Chemicals.** Cadmium oxide (CdO, 99.5%, trace metals basis), mercury chloride ($HgCl_2$, ≥99.5%), zinc acetate (Zn(ac)$_2$, 99.99%, trace metals basis), manganese (II) acetate tetrahydrate (Mn(ac)$_2$·4H$_2$O, ≥ 99.99%, trace metals basis), selenium (Se, ≥ 99.99%, trace metals basis), sulfur (99.98%, trace metals basis), oleic acid (OA, 90%, technical grade), oleylamine (OAm, ≥ 98%) 1-octadecene (ODE, 90%, technical grade), octadecylphosphonic acid (ODPA, 97%), trioctylphosphine oxide (TOPO, 99%), diphenylphosphine (DPP, 98%), tributylphosphine (TBP, 95%), lauryl methacrylate (96%), ethylene glycol dimethacrylate (98%), diphenyl (2,4,6-trimethylbenzoyl) phosphine oxide (97%), 1,2-dichloroethane (DCE, anhydrous, 99.8%), nitric acid (70%), toluene (anhydrous, 99.8%), ethanol (EtOH, anhydrous, 99.5%), 2-propanol (≥ 99.5%), and acetone (≥ 99.5%) were purchased from Sigma Aldrich. Trioctylphosphine (TOP, 97%) was purchased from Strem Chemicals. IR-26 dye (photoluminescence quantum efficiency of 0.05%) was purchased from BOC Science. All chemicals were used as received without further purification.

**Precursor preparation.** Stock solutions of TOP-Se (0.5 M), ODE-S (0.1 M), TOP-DPP-Se (TBP-ODE-Se, and Zn precursor were prepared before the QD synthesis. For TOP-Se, Se powder (0.395 g) and TOP (10 mL) were loaded in a three-neck flask in a nitrogen-filled glove box, placed on the Schlenk line, and degassed under vacuum at room temperature for 30 min. The flask was heated to 250 °C under $N_2$ gas and the temperature was maintained for 1 h. For ODE-S, sulfur powder (32 mg) and ODE (10 mL) were mixed by sonication until the solution became transparent. For TOP-DPP-Se, TOP-DPP solution was first prepared by stirring a mixture of TOP (9 mL) and DPP (45 $\mu$L) in the glove box. Se powder (850 mg) was then mixed with TOP-DPP solution (3.6 mL) and stirred overnight at 100 °C. For TBP-ODE-Se, Se powder (580 mg) in TBP (5 mL) and ODE (5 mL) mixture were stirred overnight at 100 °C in the glove box. For Zn precursor solution, a mixture of $Zn(ac)_2$ (0.392 g), TOPO (1.0 g), ODE (16 mL), and TOP (4 mL) was heated to 200 °C with a $N_2$ purging and the reaction proceeded for 1 h. All precursor solutions were stored in the nitrogen-filled glove box.

**CdSe QD synthesis.** CdO (60 mg), ODPA (280 mg), and TOPO (3.0 g) were loaded in a 50 mL three-neck flask, placed on the Schlenk line, and degassed under vacuum for 1 h at 150 °C. The reaction mixture was heated to 350 °C under a $N_2$ atmosphere and the temperature was maintained for 5 h to obtain a transparent solution. A mixture of TOP-DPP solution (1.8 mL) was added at 350 °C. The solution was further heated to 380 °C and TOP-DPP-Se (0.43 mL) was quickly injected. The flask was removed from the heating mantle after 3 min and allowed to cool to ~60 °C.

Toluene (10 mL) was added for dispersion and the QD solution was stored in ambient air. For the use of CdSe QDs, the solution was isolated by centrifugation with EtOH. The precipitated QDs were redispersed in toluene and then collected by centrifugation with EtOH, and this step was repeated two more times. The desired mass of QDs was obtained after the precipitative washing steps.

**Mn diffusion doping procedure.** Mn(ac)$_2$·4H$_2$O (100 mg), ODE (20 mL), OAm (4 mL), and OA (4 mL) were loaded in a 100 mL three-neck flask, placed on the Schlenk line, and degassed under vacuum for 1 h at 110 °C. The reaction mixture was heated to 270 °C under an N$_2$ atmosphere and maintained for 5 min. To remove the generated vapor byproducts, the reaction mixture was cooled to 110 °C and degassed for 1 h. The temperature was increased to 300 °C with an N$_2$ purging and CdSe QDs (20 mg) in TBP-ODE-Se (1 mL) were injected. The reaction proceeded for 15 min and the flask was allowed to cool to room temperature by removing the heating mantle. The solution was split between two conical tubes and isolated by centrifugation with EtOH. The products were purified again by centrifugation with toluene and EtOH.

**CdSe/HgSe/ZnS and Mn:CdSe/HgSe/ZnS QD synthesis**. The CdSe/HgSe/ZnS and Mn:CdSe/HgSe/ZnS QDs were synthesized using undoped CdSe and Mn-doped CdSe QDs, respectively. HgCl$_2$ (0.1 g) and OAm (5 mL) were loaded in a 50 mL three-neck flask, placed on the Schlenk line, and degassed under vacuum for 1 h at 110 °C. The temperature was reduced to 90 °C, and CdSe (20 mg) or Mn-doped CdSe (20 mg) in ODE (1 mL) was injected under an N$_2$ atmosphere. After 1 min, TOP-Se (0.5 M, 0.4 mL) in ODE (2.6 mL) was added at a rate of 0.5 mL/min using a syringe pump. The QDs were isolated by centrifugation with EtOH and purified again by centrifugation with toluene and EtOH. The products were redispersed in ODE (5 mL) and a Zn precursor solution (0.2 mL) and degassed under vacuum using the Schlenk line for 1 h at room temperature. The solution was heated to 240 °C at an average heating rate of 10 °C/min, during which a mixture of ODE-S (0.1 M, 0.5 mL) and Zn precursor solution (1 mL) was injected from 100 °C at a rate of 0.1 mL/min using a syringe pump. The reaction was continued for another 5 min after the addition of the Zn and S precursor solution, after which the flask was allowed to cool to room temperature by removing the heating mantle. The products were precipitated with EtOH and toluene dispersions and isolated by centrifugation. The precipitative washing steps were performed two more times.

## Material characterization

**TEM imaging.** TEM images of QDs deposited onto continuous carbon-coated Cu grids (Electron Microscopy Science) were obtained with a JEOL 2010F microscope operated at 200 kV. For TEM studies, the QD samples were purified by centrifugation with toluene and ethanol two times and redispersed in toluene. The purified QDs were drop-cast onto a TEM grid from diluted toluene dispersions and dried under vacuum overnight.

**Elemental analysis.** To determine Mn content in the Mn-doped QDs, the QD samples were studied using inductively coupled plasma atomic emission spectroscopy (ICP-AES) implemented on an ICPE-9000 (Shimadazu) spectrometer. For the ICP-AES measurements, the QDs were thoroughly purified by centrifugation with toluene and ethanol to minimize surface-bound Mn. The purified QDs were then digested in a 1% nitric acid solution in deionized water.

**Optical absorption spectroscopy.** Optical absorption spectra were recorded using a Perkin-Elmer Lambda 950 ultraviolet-visible-near infrared (NIR) spectrophotometer using QD solutions loaded into a 1 mm path length quartz cuvette. Samples were prepared by dispersing QDs in anhydrous

TCE to avoid NIR absorption by solvent C-H vibrations. Solvent-related background was subtracted from the measured spectra.

**Steady-state PL spectra**. For PL measurements, QDs dispersed in anhydrous TCE were loaded into a quartz cuvette with a path length of 1 mm and continuously stirred during the measurements to avoid photocharging. QDs were excited at 808 nm using a laser diode (IS808-100, MeshTel). The laser beam was mechanically chopped at 400 Hz using a 3501 optical chopper controller (New Focus). NIR PL spectra were obtained using a liquid-nitrogen-cooled InSb detector (IS-2.0, InfraRed Associates Inc.) equipped with a grating monochromator (Acton SP 2300). The detector was coupled to a preamplifier (INSB-1000, Infrared Systems Development) followed by an SR 830 lock-in amplifier (Standford Research Systems). The PL quantum yields of the NIR emitting QDs were quantified using the organic dye IR-26 (dissolved in DCE) as a standard. According to the specifications, the PL quantum yield of the dye was 0.05%. For visible PL measurements, the QD were dispersed in toluene. Visible PL spectra were recorded using a Horiba Scientific FluoroMax+ spectrofluorimeter with excitation at 450 nm.

**Magnetic circular dichroism (MCD) measurements.** For MCD measurements, QDs were embedded in a polymer matrix. A sample of Mn:CdSe/HgSe/ZnS QDs (~5 mg) was mixed with diphenyl (2,4,6-trimethylbenzoyl) phosphine oxide (1 mg) and lauryl methacrylate (100 μL). The solution was placed between two glass slides separated by a spacer (made from Goretex tape) and held together with clamps. To initiate polymerization, the sample was exposed to 365 nm light from a UV lamp for 30 min.

The QD film was mounted in a variable temperature insert of a superconducting magnet with direct optical access. Spectrally narrow probe light with tunable wavelength was obtained from a halogen lamp. The probe light was mechanically chopped at 95 Hz and modulated between left and right circular polarizations by a photoelastic modulator (PEM) at 50 Hz. The probe beam was focused through the QD sample and detected by a temperature-compensated, variable-gain InGaAs avalanche photodetector. The MCD signal ($\Delta A_{MCD}$) was obtained from the ratio of the difference and sum of the transmitted probe light intensities measured for two opposite circular polarizations using two separate lock-in amplifiers. One was locked to the low-frequency mechanical chopper (sum signal) and the other to the PEM (difference signal). The MCD signal obtained with this approach is proportional to the Zeeman splitting between spin-dependent light-absorbing transitions. The MCD measurements were performed in a magnetic field of up to 6 T applied along the probe propagation path (Faraday configuration) at different sample temperatures.

**Transient absorption measurements.** The purified QDs in TCE were loaded into a quartz cuvette with a path length of 1 mm and continuously stirred to avoid photocharging. TA measurements were performed using a pump-probe setup based on a regeneratively amplified femtosecond Yb:KGW laser (Pharos, Light Conversion) generating 190 femtosecond pulses at 1030 nm with a repetition rate of 500 Hz. To obtain a tunable pump wavelength, approximately half of the fundamental laser output was used to seed a high harmonic generator (HIRO HHG, Light Conversion) to produce the second (515 nm) or third (343 nm) harmonic radiation. Alternatively, it was used to pump an optical parametric amplifier (OPA, Orpheus, Light Conversion), followed by a harmonic generator (LYRA, Light Conversion).

After wavelength conversion, the pump beam of the desired wavelength was modulated at 1 kHz using an optical chopper synchronized to select every other pulse from the pulse train. The pump pulses were focused onto the sample to a spot of 100–150 μm in diameter. The remaining half of the main laser output at 1030 nm was fed into a 4 ns optical delay line and tightly focused onto a YAG crystal (EKSMA Optics) to generate a broadband white light continuum. The generated white light was focused onto the sample to a spot of less than 90 μm in diameter at the center of the focused pump beam. The pump and probe beam sizes were measured at the overlap location using a beam profiler (DataRay).

Differential absorption spectra were obtained by scanning the pump-probe delay and recording the spectra of transmitted white light pulses with and without the pump pulse using an InGaAs spectrophotometer (AvaSpec-NIR 512-1.7 TEC, Avantes). Correction for probe 'chirp' (spectro-temporal broadening) was performed by finding the zero pump-probe delay time ($t_{0,\lambda}$) at each wavelength ($\lambda$). The resulting relationship was fitted to $t_{0,\lambda} = Ae^{-\lambda/\lambda_0} + B$, where $A$, $B$ and $\lambda_0$ were the fitting parameters. The measured TA data were corrected for the 'chirp' by shifting 'time zero' at each wavelength by $t_{0,\lambda}$.

The average nominal per-dot excitonic occupancy generated by the pump pulse, $\langle N \rangle$, was found from $\langle N \rangle = (\sigma P_{opt})/(h\nu_p A f)$, where $\sigma$ is the QD absorption cross-section at the pump wavelength, $P_{opt}$ is the optical power of the pump beam, $A$ is the excited spot area, $f$ is the pump pulse repetition rate, and $h\nu_p$ was the pump photon energy. The QD absorption cross-section, was determined from the dependence of a TA signal at a long pump-probe delay of ~1 ns on the pump fluence $j_p = P_{opt}/(Af)$. At long delays, when all multiexcitons have already decayed to form single excitons, the band-edge (1S) absorption bleaching ($\Delta\alpha_{1S,long}$) is proportional to the total number of initially excited QDs ($p_{exc}$), $|\Delta\alpha_{1S,long}| = Cp_{exc}$.

In the case of Poisson statistics of photon absorption events, $p_{exc} = 1 - p_0$, where $p_0 = e^{-\langle N \rangle}$ is the fraction of unexcited QDs in the QD ensemble. $\langle N \rangle$ can be related to $\sigma$ by $\langle N \rangle = \sigma j_p (h\nu_p)^{-1}$, which gives $|\Delta\alpha_{1S,long}| = C[1 - \exp(-\sigma j_p (h\nu_p)^{-1})]$. This expression was used to fit the measured dependence of $|\Delta\alpha_{1S,long}|$ on $j_p$ with $C$ and $\sigma$ as fitting parameters. Typically, this procedure was applied to determine the absorption cross-section for $h\nu_p$ = 1.2 eV. The absorption cross-sections at other wavelengths were obtained by scaling $\sigma_{1.2\,eV}$ according to the linear absorption coefficient spectrum of the QD sample.

**Photocurrent measurements.** The photoconductor devices were prepared by photolithography using a Suss MicroTec MJB3 mask aligner and a Varian electron beam evaporator. Elemental Cr and Au were deposited on a boron-doped p-type silicon wafer at a rate of 1 nm/s. The fabricated devices contained 48 interdigitated fingers forming 23 channels. The fingers were 205 nm high (5 nm Cr and 200 nm Au), 20 μm wide, and 990 μm long. Adjacent fingers were separated by a gap of 10 $\mu$m.

To remove long-chain ligands, reference (undoped) and Mn-doped QD samples were precipitated with EtOH and toluene dispersions and isolated by centrifugation. The precipitative washing steps were performed five times. To form a densely packed film, the purified QDs were dispersed in toluene (30 mg/mL, 3$\mu$L) and drop-cast onto the pre-patterned substrate, followed by a short drying period (~10 min). The deposition/drying steps were repeated two more times. The final QD film was left to dry at room temperature in ambient atmosphere overnight. The thickness of the QD layer obtained in this way was approximately 100 nm.

Fiber-coupled light-emitting diodes (LEDs) M1050F3, M880F2, M740F2, M617F2, M530F3, and M420F2 (Thorlabs) were used to generate light with wavelengths of 1050, 880, 740, 617, 530, and 420 nm, respectively. The LED emission was focused onto the sample using an optical fiber and a lens. The beam diameter at the sample location, measured using a beam profiler, was 3.5 mm. The incident power was controlled using a continuously variable neutral density filter (NDC-100C-4M, Thorlabs). Dark current and photocurrent measurements were performed using a CHI660B electrochemical workstation (CH Instruments) using the linear sweep voltammetry method.

**Data availability.** All data that support the findings of this study are available from the corresponding author upon request.

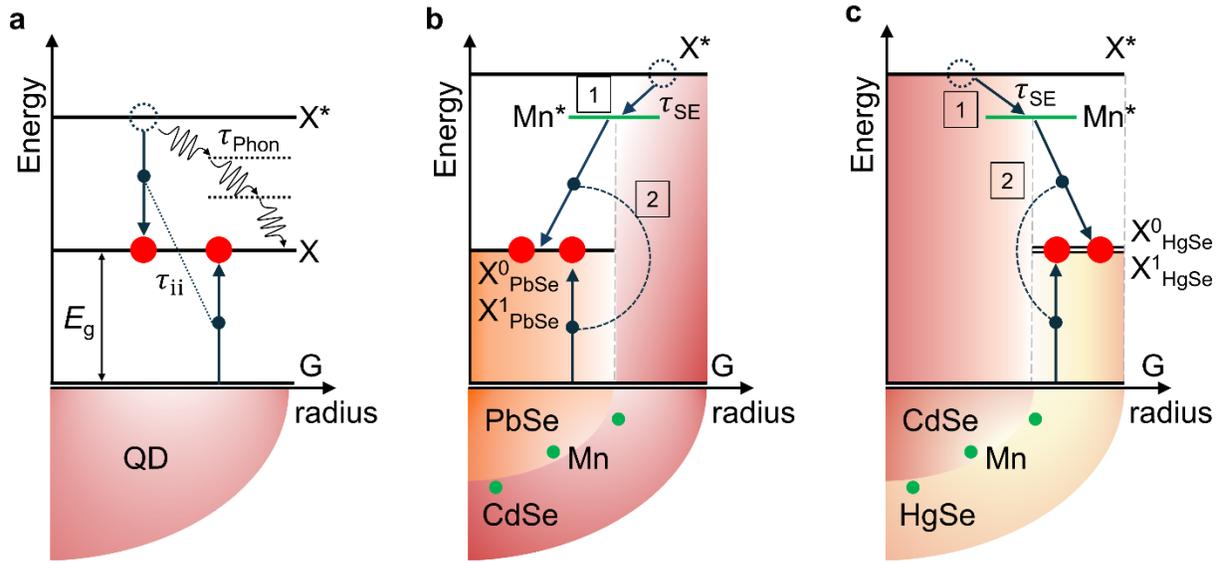

**Fig. 1 | Conventional CM and SE-CM. a,** In conventional CM, impact ionization, due to which an additional electron-hole pair (extra exciton) is generated, competes with phonon-mediated cooling of the hot exciton (X*). **b**, SE CM in Mn-doped PbSe/CdSe QDs occurs via two SE steps. Step 1 is the SE energy transfer from the hot exciton the Mn ion. In step 2, the excited Mn ion (Mn*) relaxes to form a biexciton in the PbSe core. Due to the spin-conservation requirement, the biexciton consists of a bright exciton ($X^0_{PbSe}$, spin 0) and a dark exciton ($X^1_{PbSe}$, spin 1). **c**, In Mn-doped 'inverted' CdSe/HgSe QDs, SE-CM occurs via two similar SE steps. Importantly, the two final excitons generated by this process ($X^0_{HgSe}$ and $X^1_{HgSe}$) are located in the HgSe shell. This makes these two excitons assessable electrically, an advantage over PbSe/CdSe QDs for which carrier extraction is complicated by the potential barrier created by the CdSe shell.

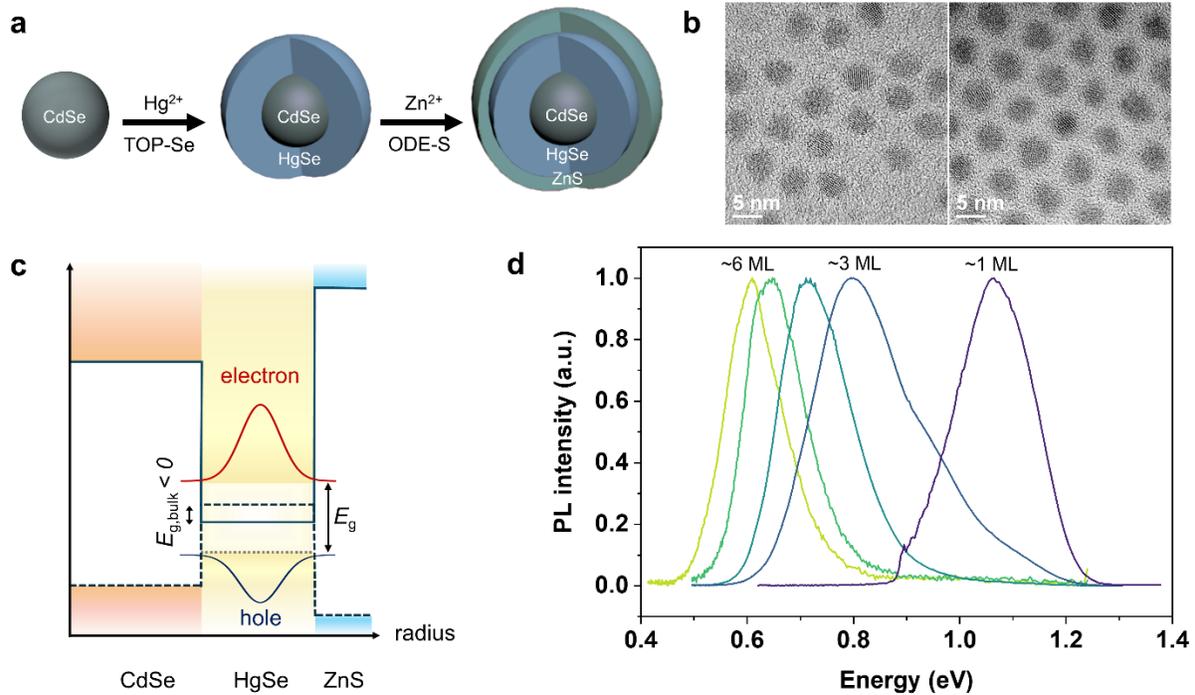

**Fig. 2 | Synthesis of undoped CdSe/HgSe QDs with controlled shell thickness. a,** Schematic representation of the synthetic procedure used to prepare CdSe/HgSe QDs with a final ZnS protective layer. The prefabricated CdSe cores react with $Hg^{2+}$ (derived from $HgCl_2$) and TOP-Se (trioctylphosphine selenium) leading to the epitaxial growth of the HgSe shell. The thickness of the HgSe shell can be controlled by the amount of the reagents and the reaction temperature. The protective layer of ZnS is prepared by adding $Zn^{2+}$ (derived from zinc acetate) and ODE-S (octadecene-sulfur) to the reaction at elevated temperature (240 °C). **b,** TEM images of CdSe cores (left, core radius is 2.3 nm) and CdSe/HgSe QDs with an approximately 1 monolayer thick HgSe shell (right). **c,** Approximate band diagram of CdSe/HgSe/ZnS QDs. The band edge transition energy (the QD bandgap) is determined primarily by the thickness of the HgSe shell since the wavefunctions of both the electron and the hole (red and blue lines, respectively) are predominantly shell localized. **d,** PL spectra of a series of CdSe/HgSe/ZnS QD samples with a fixed CdSe core radius ($r_{CdSe}$ = 2.3 nm) and $H_{HgSe}$ varying from 0.3 to 1.9 nm (~1 to ~6 HgSe MLs). The ZnS layer thickness is ~1 ML. As expected, based on the band diagram in **c**, the PL peak position shows a strong dependence on $H_{HgSe}$.

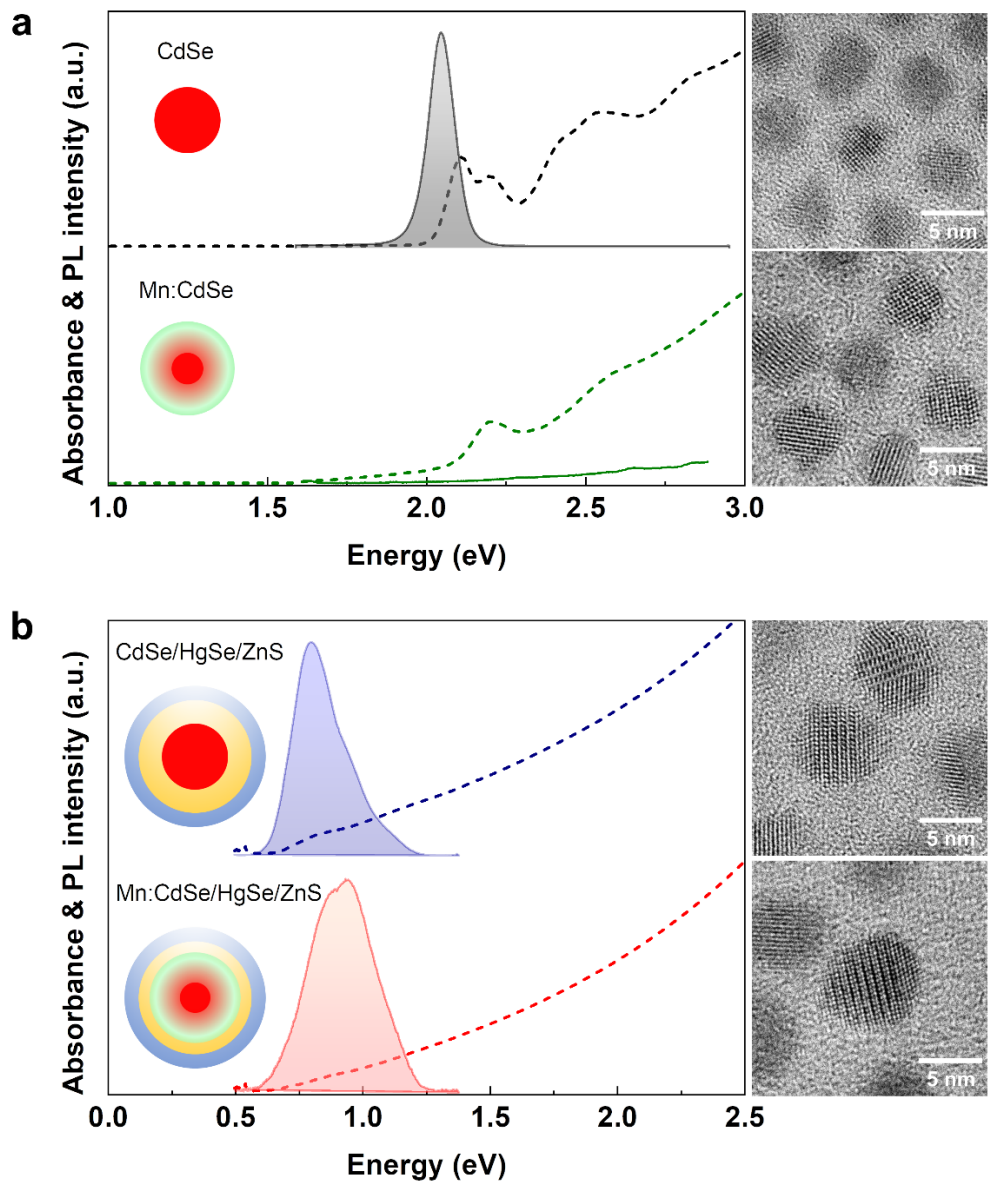

**Fig. 3 | Structural characteristics and optical spectra of undoped and Mn-doped CdSe and CdSe/HgSe/ZnS QDs. a,** Optical absorption (black and green dashed lines) and PL (grey shaded profile) spectra (left) of CdSe (top) and Mn:CdSe (bottom) QDs, together with their TEM images (right). The duration of the doping procedure was 15 minutes. After Mn doping, the QD radius increases from 2.3 to 2.4 nm, while the 1S absorption peak blueshifts, corresponding to incorporation of Mn ions into the CdSe cores. The PL emission observed for the original CdSe QDs is quenched after doping, due to the surface defects introduced during the doping procedure. **c,** The same information for the CdSe/HgSe/ZnS (top) and Mn:CdSe/HgSe/ZnS (bottom) QDs prepared from the CdSe cores shown in **a**. The HgSe and ZnS layer thicknesses are ~1.0 and ~0.4 nm, respectively. After the deposition of the HgSe shell, both the absorption (blue and red dashed lines) and PL (blue and red shaded profiles) spectra of the QDs are shifted to the near-infrared energy range. The insets are schematic images of undoped and Mn-doped CdSe and CdSe/HgSe/ZnS QDs. Segments of QDs made of CdSe, HgSe. and ZnS are shown in red, yellow, and blue, respectively. The QD region containing Mn ions is shown by green shading.

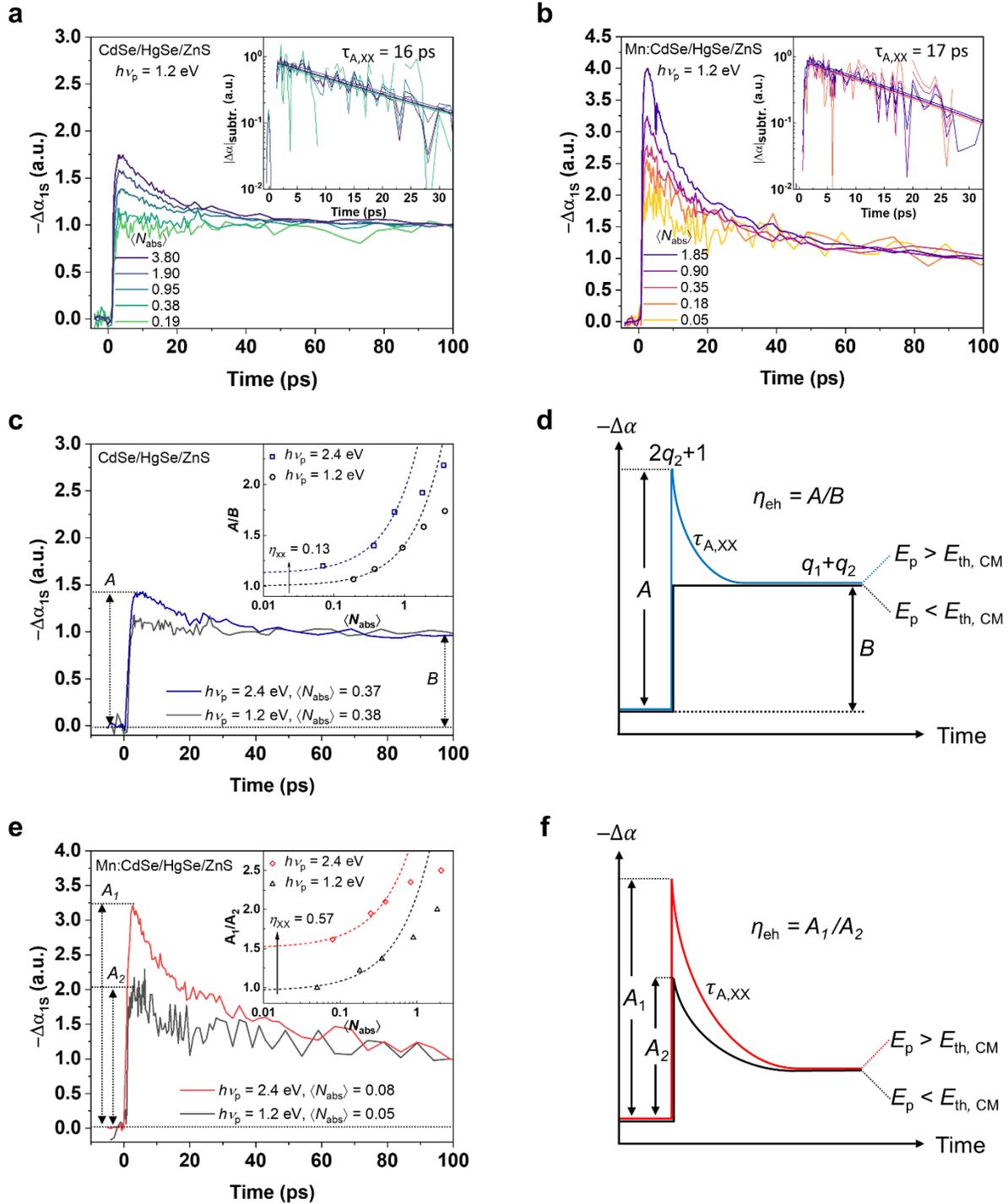

**Fig. 4 | TA measurements of CM in CdSe/HgSe/ZnS and Mn:CdSe/HgSe/ZnS QDs. a,** Pump-intensity dependent 1S bleaching dynamics of reference (undoped) CdSe/HgSe/ZnS QDs obtained using 1.2-eV excitation (without CM) with variable pump fluence for which $\langle N_{abs} \rangle$ changes from 0.19 to 3.80. The inset shows the biexciton dynamics isolated by subtracting the tail-normalized TA trace obtained at low pump fluence ($\langle N_{abs} \rangle = 0.19$) from the TA traces for higher pump fluences. Based on a single-exponential fit, the biexciton lifetime is 16 ps. **b,** A similar data set for Mn-

doped CdSe/HgSe/ZnS QDs measured under the same conditions. Based on the extracted initial fast dynamics (inset), the biexciton lifetime is 17 ps. **c,** 1S bleaching dynamics of undoped QDs obtained using 1.2 eV (black) and 2.4 eV (blue) excitation with $\langle N_{abs} \rangle$ = 0.38 and 0.37, respectively. The inset shows the *A/B* ratio as a function of $\langle N_{abs} \rangle$ for the 2.4 eV (blue symbols) and 1.2 eV (black symbols) excitation. Extrapolating a linear fit of the 2.4 eV data (blue line) to $\langle N_{abs} \rangle$ = 0 yields $\eta_{eh}$ =1.13 and $\eta_{XX}$ = 0.13. A similar fit of the 1.2 eV data yields $\eta_{eh}$ =1 (black line). **d,** Illustration of the determination of the CM efficiency from the 1S bleaching dynamics measured for $\langle N_{abs} \rangle \ll 1$. In the regime when the CM does not produce more than 1 extra exciton, $\eta_{eh}$ can be determined from the ratio of the TA signals before Auger decay (*A*, early time TA peak amplitude) and after completion of Auger recombination (*B*, slow-relaxing single-exciton background). **e,** 1S bleaching dynamics of the Mn-doped QDs obtained using excitation with 1.2 eV (black) and 2.4 eV (red) photons with $\langle N_{abs} \rangle$ = 0.05 and 0.08, respectively. The inset shows a plot of the $A_1/A_2$ ratio versus $\langle N_{abs} \rangle$ for the 1.2 eV (black symbols) and 2.4 eV (red symbols) excitation. Based on a linear extrapolation of the 2.4-eV data, $\eta_{eh}$ = 1.57 and $\eta_{XX}$ = 0.57. **f,** Illustration of a modified procedure for determining $\eta_{eh}$ in the presence of fast single-exciton decay.

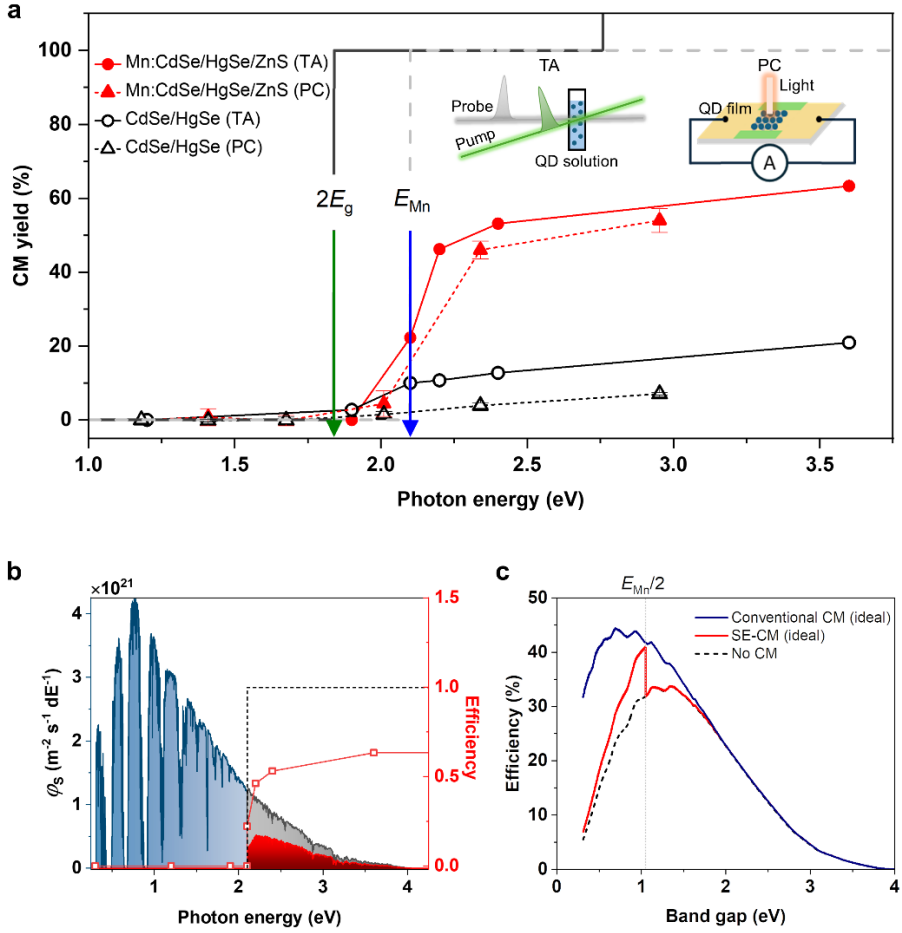

**Fig. 5 | CM yields obtained from TA and photocurrent measurements and implications of SE-CM for solar energy conversion. a,** CM yields for CdSe/HgSe/ZnS and Mn:CdSe/HgSe/ZnS QDs obtained by TA spectroscopy (black and red circles, respectively) and photocurrent (PC) measurements (black and red triangles, respectively) as a function of pump photon energy. The data sets obtained using the two methods (shown in the insets) are in excellent agreement and indicate significantly higher CM yields for Mn-doped QDs compared to undoped QDs. The 'ideal' CM yields for conventional CM ($\eta_{CM}$ increases by 100% for each increment in photon energy by $E_g$ for energies above $2E_g$) and SE-CM ($\eta_{CM}$ = 100% for photon energies above $E_{Mn}$) are shown by solid black and dashed grey lines, respectively. **b,** The spectrum shaded blue shows the spectral distribution of the solar photon flux as a function of photon energy. The 'ideal' SE-CM efficiency with CM onset at $E_{Mn}$ =2.1 eV is shown as a dashed black line. The corresponding flux gain is indicated by gray shading. The experimentally measured SE-CM efficiencies are plotted as red squares while, and their linear interpolation is shown as a solid red line. The corresponding solar flux enhancement is shown by red shading. **c,** Simulated PCE plotted as a function of the material bandgap for the cases without CM (dashed black line), with 'ideal' conventional CM (solid blue line), and with 'ideal' SE-CM (solid red line). In the no-CM case, $\eta_{PCE}$ reaches 33.6% at the optimal bandgap of 1.33 eV. In the case of 'ideal' conventional CM, the maximum PCE is 44.4% and is realized at $E_g$ = 0.69 eV. In the 'ideal' SE-CM case, PCE reaches a maximum value of 41.0% at $E_g$ = $E_{Mn}/2$ = 1.05 eV.

Supplementary Information for:

# Photocurrent Enhancement due to Spin-Exchange Carrier Multiplication in Films of Manganese-Doped 'Inverted' CdSe/HgSe Quantum Dots

Jungchul Noh, Clément Livache, Donghyo Hahm, Valerio Pinchetti, Ho Jin, Changjo Kim, and Victor I. Klimov*

Nanotechnology and Advanced Spectroscopy Team, C-PCS, Chemistry Division, Los Alamos National Laboratory, Los Alamos, NM 87545, USA

*klimov@lanl.gov

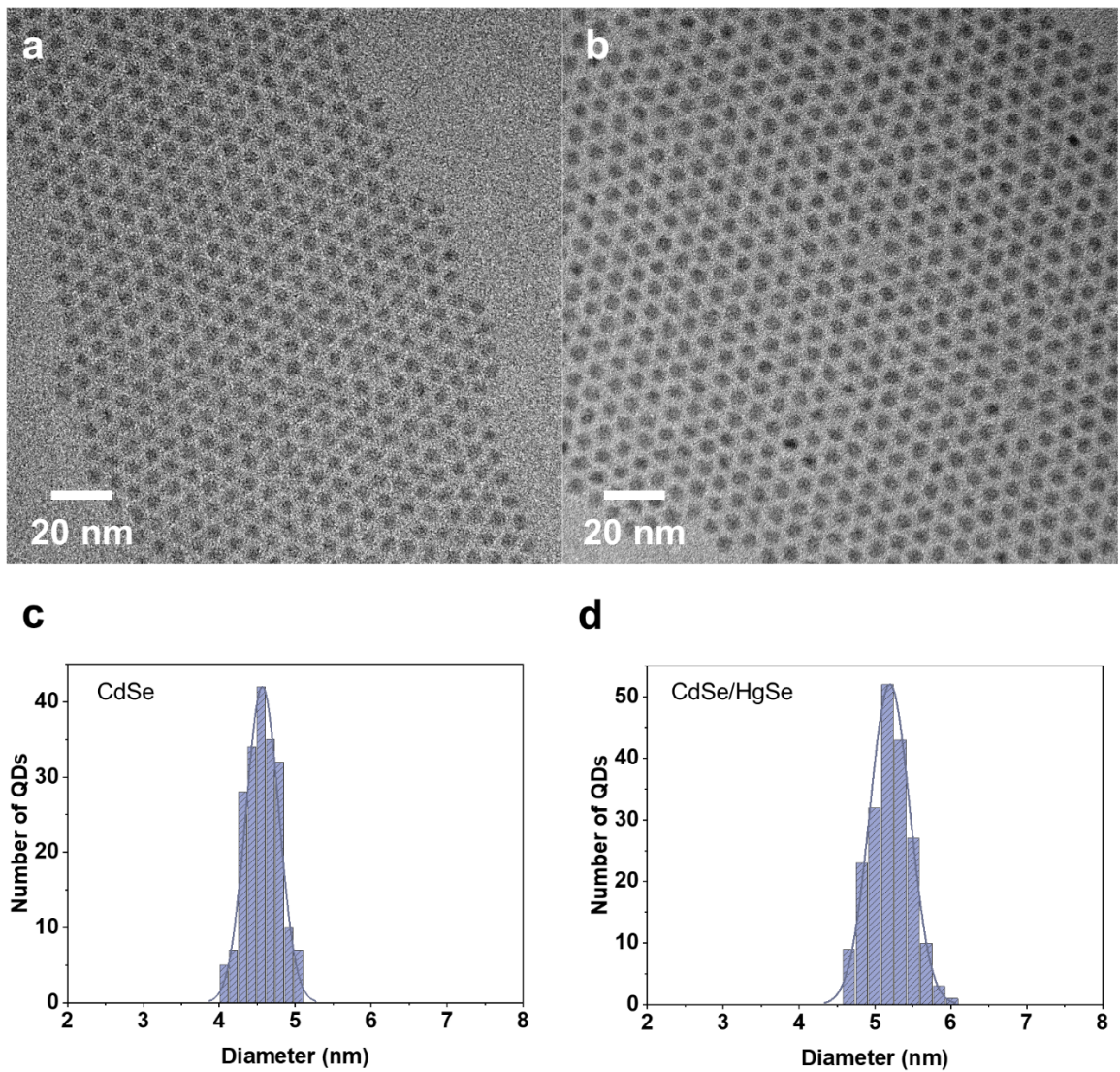

**Supplementary Fig. 1 | Transmission electron microscopy (TEM) images of CdSe and CdSe/HgSe quantum dots (QDs).** TEM images of CdSe (a) and core/shell CdSe/HgSe (b) QDs. (**c,d**) QD size histograms obtained by analyzing the dimensions of 200 particles. The average diameters of CdSe and CdSe/HgSe QDs are 4.6 $\pm$ 0.22 and 5.2 nm $\pm$ 0.26 nm, respectively. Both samples are highly monodisperse with a standard deviation of size of only 5%. Based on the average diameters of core and core/shell particles, the shell thickness is 0.3 nm, corresponding to ~1 HgSe monolayer.

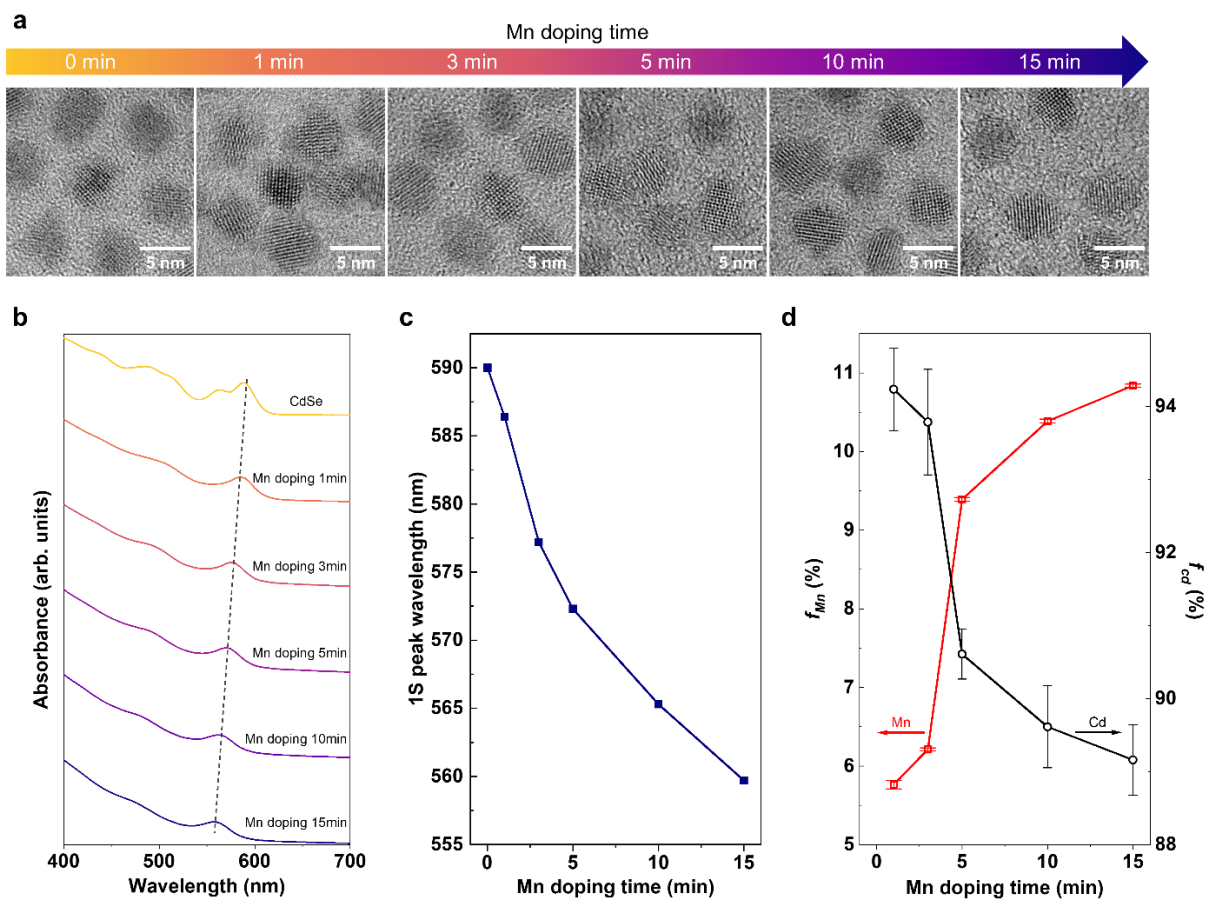

**Supplementary Fig. 2 | Diffusion doping of CdSe QDs with manganese (Mn). a**, TEM images of CdSe QDs and Mn-doped CdSe QDs prepared using different diffusion doping reaction times ($t_{Mn}$ varied from 1 to 15 min). The original diameter of CdSe QDs is 4.6 nm. After the 15-min doping procedure, the QD diameter increases to ~4.8 nm. **b,** Absorption spectra of undoped and Mn-doped CdSe QDs. The band-edge (1S) absorption peak blueshifts with the increase of $t_{Mn}$, indicating the increasing amount of Mn incorporated into the QDs. **c,** Position of the 1S peak as a function of $t_{Mn}$. **d,** Relative fractions of Mn and Cd ions in Mn-doped CdSe QDs ($f_{Mn}$ and $f_{Cd}$, respectively) as a function of the doping reaction time. These data were obtained by elemental analysis using inductively coupled plasma atomic emission spectroscopy (ICP-AES). For each sample, ICP-AES measurements were repeated three times. The mean values are shown as symbols and the standard deviation as error bars (see Supplementary Table 1).

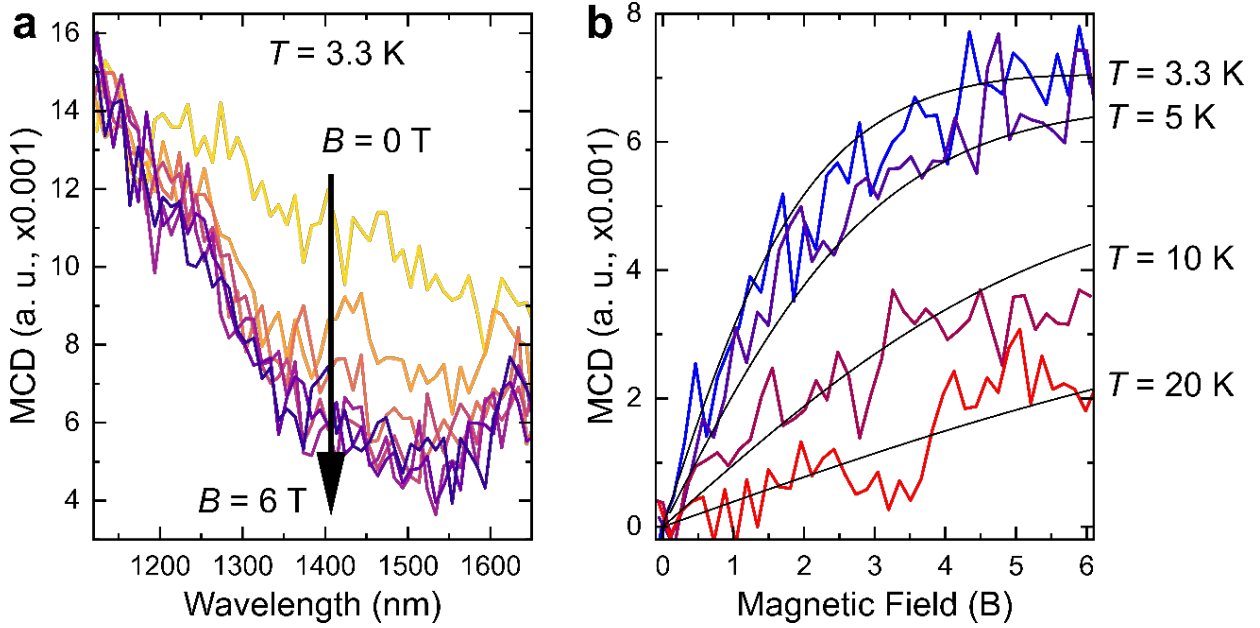

**Supplementary Fig. 3 | Magnetic circular dichroism (MCD) measurements of Mn:CdSe/HgSe/ZnS QDs.** **a**, MCD spectra of Mn:CdSe/HgSe/ZnS QDs (sample temperature $T$ = 3.3 K) used in the studies of spin-exchange carrier multiplication (SE-CM) as a function of the magnetic field ($B$) varied from 0 to 6 T with a step of 1 T (for details of the MCD measurements, see the Methods section). **b**, The MCD signal amplitude ($\varDelta A_{MCD}$) at 1490 nm as a function of $B$ at different sample temperatures varied from 3.3 K to 20 K (colored line). The observed trends can be accurately described by the expression $\varDelta A_{MCD} = C(g_{int}\mu_B B + N_0(\alpha - \beta)x_{eff}\langle S_Z \rangle)$ (black lines), where $C$ is a proportionality constant, $g_{int}$ is the intrinsic exciton g-factor, $\mu_B$ is the Bohr magneton, $x_{eff}$ is the effective concentration of free (paramagnetic) Mn spins in the QD volume sampled by the exciton wavefunction, $\alpha$ and $\beta$ are the s-d and p-d exchange constants, respectively, $N_0$ is the number of cations per unit volume, and $\langle S_Z \rangle$ is the average projection of Mn spins along the applied magnetic field. The latter quantity can be represented as $\langle S_Z \rangle = SB_S[g_{Mn}\mu_B SB/k_B T]$, where $S$ is the spin of the Mn ion ($S$=5/2), $g_{Mn}$ is the g-factor of the Mn ion, and $B_S$ is a modified Brillouin function that depends on both $B$ and $T$ and describes the interplay between field-induced alignment (ordering) of the Mn spins and their thermally induced disordering[1]. In our analysis of the MCD data, we use literature values of exchange constants for HgSe $N_0\alpha = 0.4$ eV and $N_0\beta = -0.7$ eV (ref 1). The best agreement between the measurements and the theoretical expression for $\varDelta A_{MCD}$ is obtained with $x_{eff} = 0.8\%$.

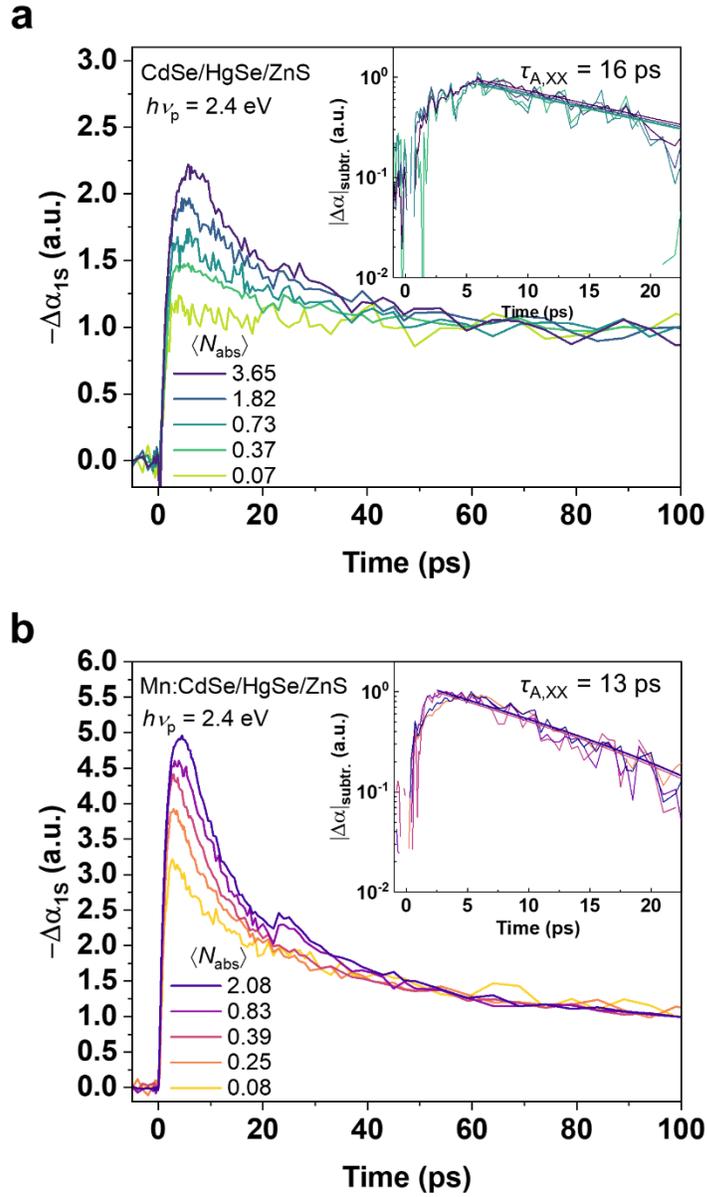

**Supplementary Fig. 4 | Band-edge (1S) bleaching dynamics of undoped and Mn-doped CdSe/HgSe/ZnS QDs obtained using excitation with 2.4 eV photons.** Pump-fluence-dependent 1S bleaching dynamics of undoped and **b,** Mn-doped (**b**) CdSe/HgSe/ZnS QDs obtained with excitation at 2.4 eV, which is above the SE-CM threshold. The pump fluence is evaluated in terms of an average number of photons absorbed per QD per pump pulse ($\langle N_{abs} \rangle$). The insets show the dynamics obtained by subtracting the lowest pump-fluence trace ($\langle N_{abs} \rangle$ = 0.07 - 0.08) from the traces at higher fluences. This procedure is used to isolate the biexciton dynamics. Based on it, the biexciton lifetimes for undoped and Mn-doped QDs are 16 and 13 ps, respectively.

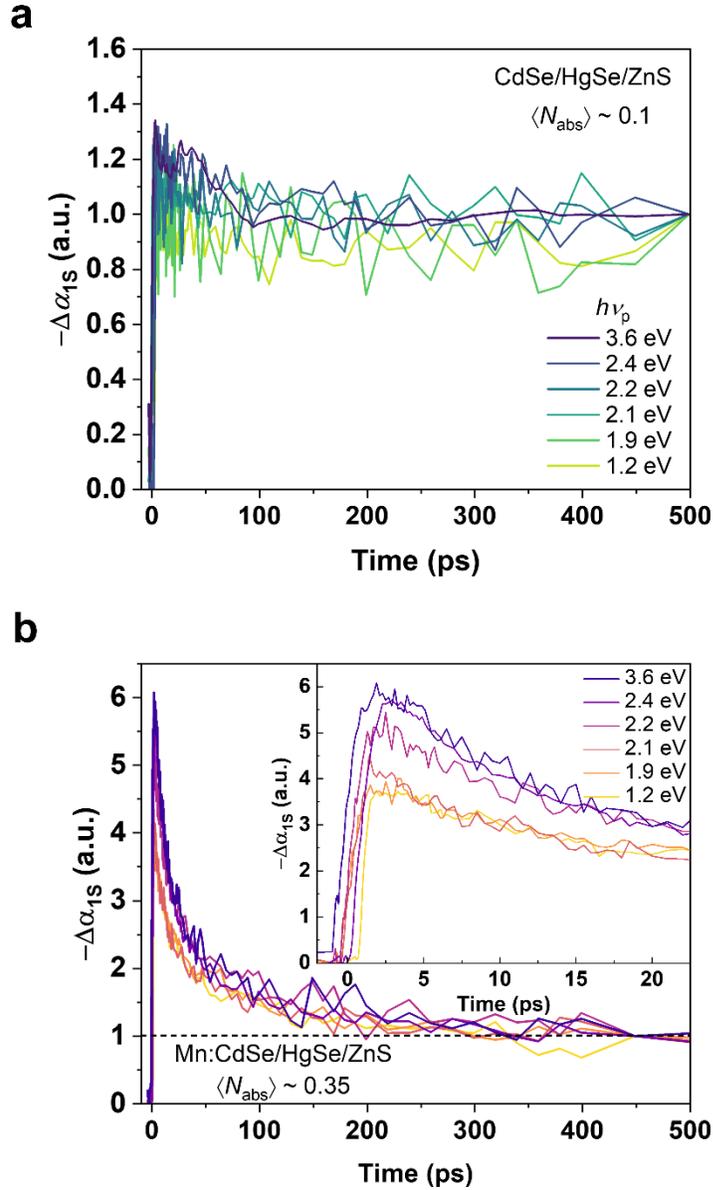

**Supplementary Fig. 5 | 1S bleaching dynamics of undoped and Mn-doped CdSe/HgSe/ZnS QDs obtained for different pump photon energies.** Tail-normalized 1S bleaching dynamics of undoped (**a**) and Mn-doped (**b**) CdSe/HgSe/ZnS QDs measured using excitation with different pump photon energies ($h\nu_p$ varied from 1.2 to 3.6 eV) at sub-single-exciton pump fluences ($\langle N_{abs} \rangle$ is ~0.1 and ~0.35 for the undoped and Mn-doped QDs, respectively). The inset in (**b**) is an expanded view of the early time 1S bleaching dynamics showing the sharp onset of SE-CM at $h\nu_p$ of ~2.1 eV, which corresponds to the Mn spin-flip transition energy ($E_{Mn}$).

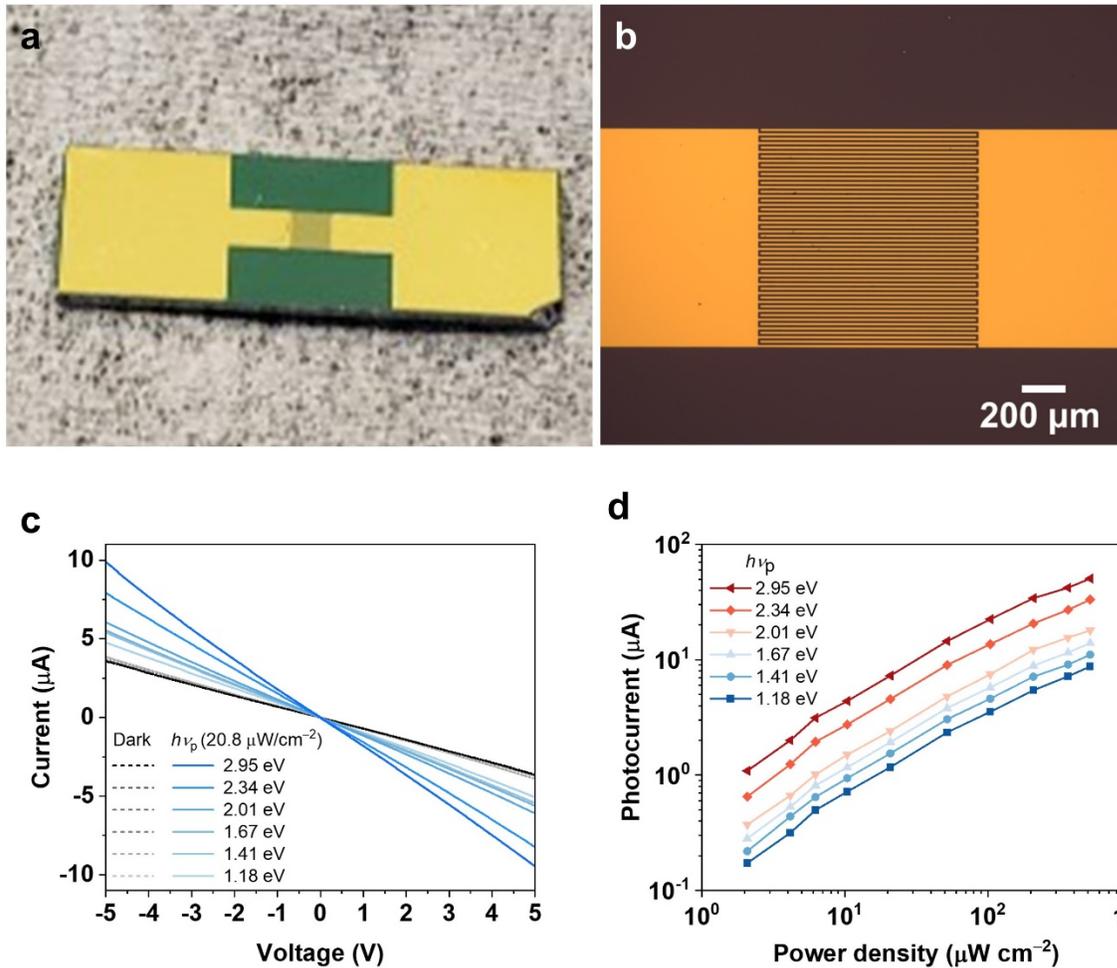

**Supplementary Fig. 6 | Photocurrent measurements of close-packed Mn-doped CdSe/HgSe/ZnS QD films. a,** Photograph (**a**) and optical microscope image (**b**) of the QD-based photoconductor device. The device consisted of 23 channels formed by 48 Cr/Au interdigitated electrodes (fingers) with the following dimensions: 205 nm height (5 nm Cr and 200 nm Au), 20 $\mu$m width, 990 $\mu$m length, and 10 $\mu$m spacing. The electrodes were fabricated on top of a Si/SiO$_2$ substrate using thermal evaporation. A QD film (~100 nm thick) was drop-cast onto the pre-patterned substrate. **c,** Current of the QD-based device measured with ($I_{light}$) and without ($I_{dark}$) illumination as a function of voltage ($V$). The photon energy of incident light was tuned from 1.18 to 2.95 eV. The excitation intensity was 20.8 $\mu$W cm$^{-2}$. **d,** Photocurrent ($I_{ph} = I_{light} - I_{dark}$) as a function of the intensity of incident light for different photon energies varied from 1.18 to 2.95 eV ($V$ = 5 V).

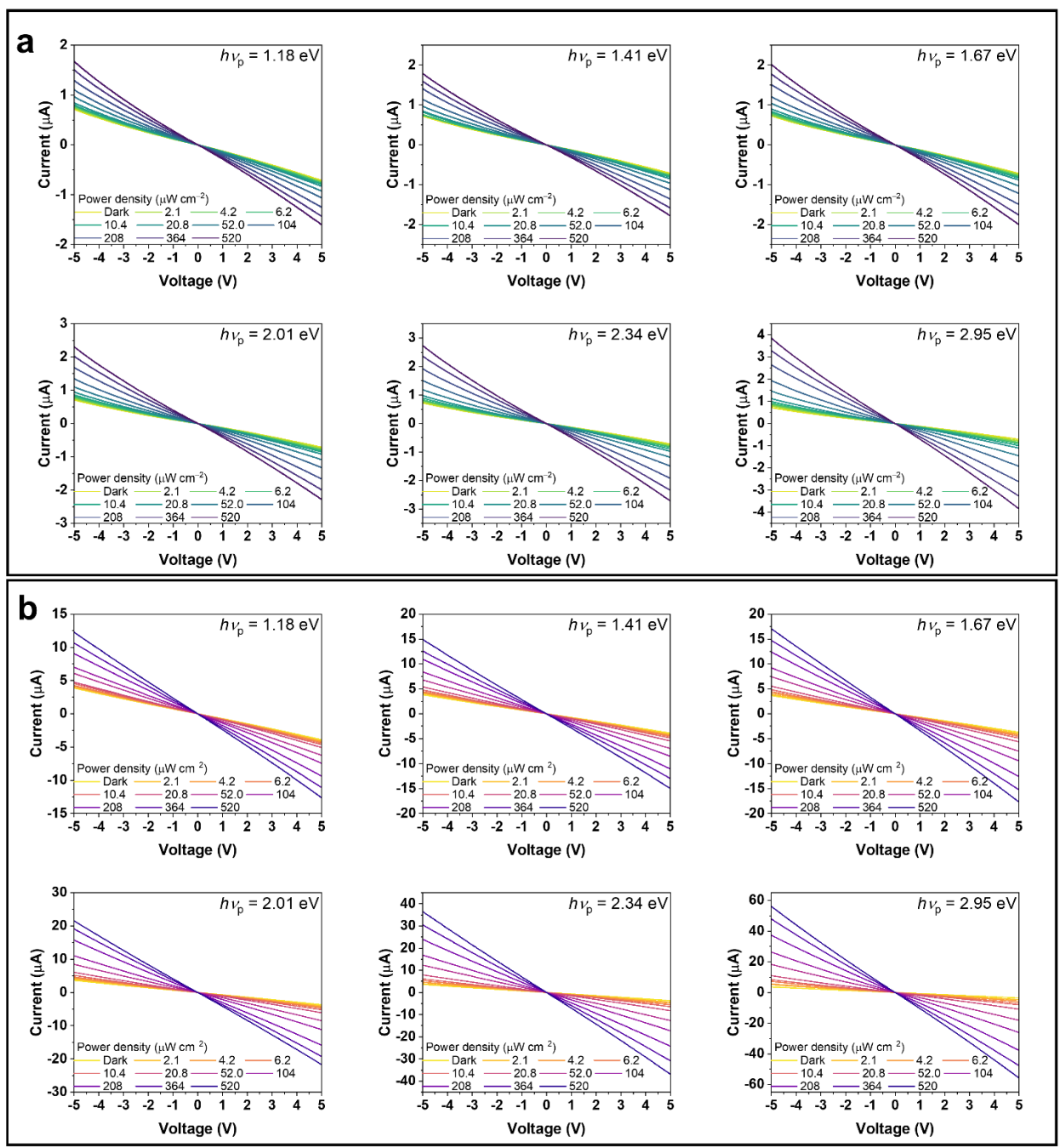

**Supplementary Fig. 7 | *I-V* characteristics of undoped (a) and Mn-doped (b) CdSe/HgSe/ZnS QD film.** *I-V* characteristics of the Mn-doped CdSe/HgSe/ZnS QDs measured with and without monochromatic illumination. In the measurements with illumination, the light intensity varied from 2.1 to 519.7 $\mu W\ cm^{-2}$. Different sub-panels correspond to different incident photon energies ranging from 1.18 to 2.95 eV.

# Supplementary Tables

**Supplementary Table 1.** Elemental analysis of Mn-doped CdSe cores and Mn-doped core/shell CdSe/HgSe/ZnS QDs.

| Sample | | | Atomic percentage (%)[a] | | | |
|---|---|---|---|---|---|---|
| | | | Cd | Mn | Hg | Zn |
| Mn:CdSe | Mn doping reaction time (min) | 1 | 94.2 | 5.8 | – | – |
| | | 3 | 93.8 | 6.2 | – | – |
| | | 5 | 90.6 | 9.4 | – | – |
| | | 10 | 89.6 | 10.4 | – | – |
| | | 15 | 89.2 | 10.8 | – | – |
| Mn:CdSe/HgSe/ZnS | | | 22.8 | 2.3 | 48.4 | 26.5 |

[a] Elemental analysis was performed by inductively coupled plasma atomic emission spectroscopy (ICP-AES) using an ICPE-9000 spectrophotometer. The values given in the table are the average of three separate ICP-AES measurements.